\documentclass[aps,pra,reprint,twocolumn,amsmath,amssymb,citeautoscript]{revtex4-1}
\usepackage{graphicx}
\usepackage{dcolumn}
\usepackage{bm}
\usepackage{latexsym,epsfig}
\usepackage{graphicx}
\usepackage[export]{adjustbox}
\usepackage{verbatim}
\usepackage{comment}
\usepackage{amsmath}
\usepackage{amssymb}
\usepackage{stmaryrd}
\usepackage{color}
\usepackage{epstopdf}
\usepackage{grffile}
\usepackage[normalem]{ulem}
\usepackage{mathtools}
\usepackage{physics}

\DeclareGraphicsExtensions{.eps}
\newcommand{\beq}{\begin{equation}}
\newcommand{\eeq}{\end{equation}}
\newcommand{\bea}{\begin{eqnarray}}
\newcommand{\eea}{\end{eqnarray}}
\newcommand{\ben}{\begin{eqnarray*}}
\newcommand{\een}{\end{eqnarray*}}
\newcommand{\bfig}{\begin{figure}}
\newcommand{\efig}{\end{figure}}

\usepackage{hyperref}
\hypersetup{
    colorlinks=true,      
    urlcolor=blue,
    citecolor=blue,
    linkcolor=blue
}

\begin{document}
\title{Interacting bosons on a Su-Schrieffer-Heeger ladder: Topological phases and Thouless pumping}
\author{Ashirbad Padhan$^{1}$, Suman Mondal$^{2}$, Smitha Vishveshwara$^{3}$ and Tapan Mishra$^{4,5,*}$}
\affiliation{$^{1}$Department of Physics, Indian Institute of Technology, Guwahati, Assam - 781039, India}
\affiliation{$^{2}$ Institute for Theoretical Physics, Georg-August-Universität Göttingen, Friedrich-Hund-Platz 1, 37077 Göttingen, Germany}
\affiliation{$^{3}$Department of Physics, University of Illinois at Urbana-Champaign, Urbana, IL USA}
\affiliation{$^4$School of Physical Sciences, National Institute of Science Education and Research, Jatni 752050, India}
\affiliation{$^5$ Homi Bhabha National Institute, Training School Complex, Anushaktinagar, Mumbai 400094, India}
 \email{mishratapan@gmail.com}
\date{\today}

\begin{abstract}
We study the topological properties of hardcore bosons on a two-leg ladder consisting of two Su-Schrieffer-Heeger (SSH) chains that are coupled via hopping and interaction. We chart out the phase diagram for the system and show that based on the relative hopping dimerization pattern along the legs, distinctly different topological phases and phase transitions can occur. When the dimerization along the legs are uniform, we find that the topological nature vanishes for even the slightest rung hopping. For staggered dimerization, the system exhibits a well defined topological character and a topological phase transition as a function of rung hopping. While the topological phase shows bond order character, the trivial phase shows the behavior of a rung-Mott insulator.
For this case, the topological nature is found to survive even in the presence of finite inter-leg interactions. Moreover, we find that the critical point of the topological phase transition shifts to a higher or a lower rung hopping strength depending on the attractive or repulsive nature of the interaction.  To highlight the marked effects of interactions, we propose a scheme involving a Thouless charge pump that provides insights for the topological phases characterized by a quantised particle transport through a periodic modulation of appropriate system parameters. In our studies, we show an interaction induced charge pumping following specific pumping protocols in the case of staggered dimerization.

\end{abstract}

\maketitle
\section{introduction}


In the past decade, a profusion of solid state, ultracold atomic, and metamaterial systems have realized topological phases of matter that beautifully conform to predictions made at the single-particle level. One of the earliest topological models proposed in the late seventies in the context of polymers,  the Su-Schrieffer-Heeger (SSH) model~\cite{ssh, ssh2}, has formed a paradigm for extensive studies. The SSH model is a tight binding model of lattice fermions having dimerized hopping and it has offered the simplest platform to realize topological phase transition in low dimensional lattice systems~\cite{Asboth2016_ssh, Batra2020}. Particularly, the SSH model exhibits a topological phase transition from one gapped phase to another as a function of the nearest neighbor hopping dimerization strength. Topological phases are characterized by zero-energy midgap states corresponding to robust, topologically protected edge bound states in finite geometries. Underlying symmetries that protect the topological character of the phases have attracted recent attention for fundamental physics as well as technological applications~\cite{Senthil_rev, Fidkowski,Xiao-Gang,Oshikawa,sptinteract, Pachos_2014, vonKlitzing2017}. The SSH model is now a frontrunner for realization in disparate experimental systems, revealing the signature edge states and distinct topological phase transitions~\cite{Atala2013, Takahashi2016pumping, Lohse2016, Mukherjee2017, Lu2014, gadway, Kitagawa2012, Leder2016}.

With well-established single-particle topological physics in place, attention has now turned to interaction effects. Here, the nature of the intrinsic particles becomes highly germane: fermions and bosons can both form insulators either due to band effects, and related topology, or due to interactions and Mott physics. Delocalized behavior is typically metal-like in fermions and superfluid in bosons. Quantum particles apart, in metamaterials, excitations could correspond to a variety of features, such as lattice distortions; deviations beyond the lowest order would give rise to effective interactions between these modes. Naturally, the SSH model has formed a key subject for understanding how interactions modify its predicted topological phases~\cite{rachel_review}. Inter-particle interactions in the SSH model and its variants are known to have significant effects on the topological character~\cite{Hayward2018, Nakagawa2018,Mondal2019,Mondal2020, fleischhauer_2013_prl, Congjun, Santos2018, Karrasch, Bermudez, Sirker_2014, ssh_bosonization, marques, Jiansheng, Yahyavi_2018}. Moreover,  interactions can even induce a topological phase and an associated phase transition~\cite{suman_v1v2,  Haldane1983, DallaTorre2006}. Interaction effects have also been generalized to two-component systems in the framework of the SSH-Hubbard model~\cite{Santos2018,mondal_sshhubbard,Yoshida2018,Nakagawa2018,Wang2015,fan2016, Yoshihito,Montorsi, Dauphin}. On the experimental front, the topological phases have been observed in a one dimensional interacting SSH model~\cite{Browaeyes2018ssh} and also in an SSH-Hubbard model~\cite{Le2020}. 
 
In this theory work, we investigate the interplay between interactions and inter-chain coupling in the case of multiple SSH chains, and its effect on topological physics. As depicted in Fig.~\ref{fig:lattice},  two-leg ladders provide a minimal setup to explore the intermediate regime between one- and two-dimensional lattices. The topological properties of these coupled SSH models have been theoretically investigated at the single particle level~\cite{song,sangmo,Li2013,smithaprb,adhip, Bercioux, Matveeva} and even realized experimentally~\cite{prodan}. The coupling between the chains, which forms the ladder, provides a degree of freedom to vary, and the possibility of new phases. In particular, if the dimerization pattern is uniform between chains (Fig.~\ref{fig:lattice}(a)), the smallest amount of rung-coupling renders topology to be weak. In that case edge states between chains can hybridize and break away from their zero-energy status.  If the dimerization pattern is staggered between chains (Fig.~\ref{fig:lattice}(b)) , such rung hopping can result in three phases~\cite{smithaprb}. Two of the phases entail having topologically protected mid-gap states, as would be found in individual chains. A third, also gapped, phase results from strong enough coupling between chains such that the mid-gap states completely disappear, and only bulk states exist. The effect of interactions on this physics has been studied in many-body spin and fermionic systems~\cite{Nersesyan,glide_reflection}. Many-body physics in bosonic SSH ladders is far from a simple extrapolation; competing effects, such as superfluidity, offer hitherto unexplored regimes and form the subject of our study.

In this paper, we thus perform an in-depth study of the topological nature of hardcore bosons on the two-leg SSH ladder in the presence of inter-leg interactions. We first review single-particle physics to set the stage for phenomena exhibited by interacting bosons. We establish how the most significant effects of topology come into play at half-filling, away from gapless superfluid phases that surround gapped phases in parameter space. On introducing inter-leg interactions, for the uniform ladder, we show how a ‘rung-Mott insulator’ phase can dictate and suppress the topological nature of the system. For the staggered ladder, starting from a topological phase at a fixed rung hopping, we show that the system sustains a robust topological phase with increasing inter-leg interactions until it reaches a critical strength at which it undergoes a phase transition to a trivial phase.  We find that the topological phase transition is sensitive to the nature of the interaction, i.e., a topological to trivial phase transition occurs at a smaller (larger) critical rung hopping when the interaction is repulsive (attractive). We do a careful study to quantify these topological properties by deriving and analyzing the excitation spectrum, finite edge polarization, quantized Berry's phase, and Thouless charge pumping properties. Our studies show that the SSH bosonic ladder has a rich range of phases, is topologically robust, and has some features that are markedly different from its single-particle and fermionic counterparts. 

The paper is organized as follows. In Sec.~\ref{sec:mm}, we discuss the model and the method used for our studies.  We obtain the single particle spectrum in Sec.~\ref{sec:single}, and employ it in Sec.~\ref{sec:many}  to derive results pertaining to the many-body hardcore bosonic system in the absence of inter-leg interactions. In Sec.~\ref{sec:inter}, we perform an involved analysis of the behavior of the SSH ladder in the presence of inter-leg interactions, concluding with a derivation and discussion of  Thouless charge pumping. Finally, in Sec.~\ref{sec:conc}, we provide a summary and outlook.

\section{Model and approach}
\label{sec:mm}
\begin{figure}[b]
\begin{center}
\includegraphics[width=1\columnwidth]{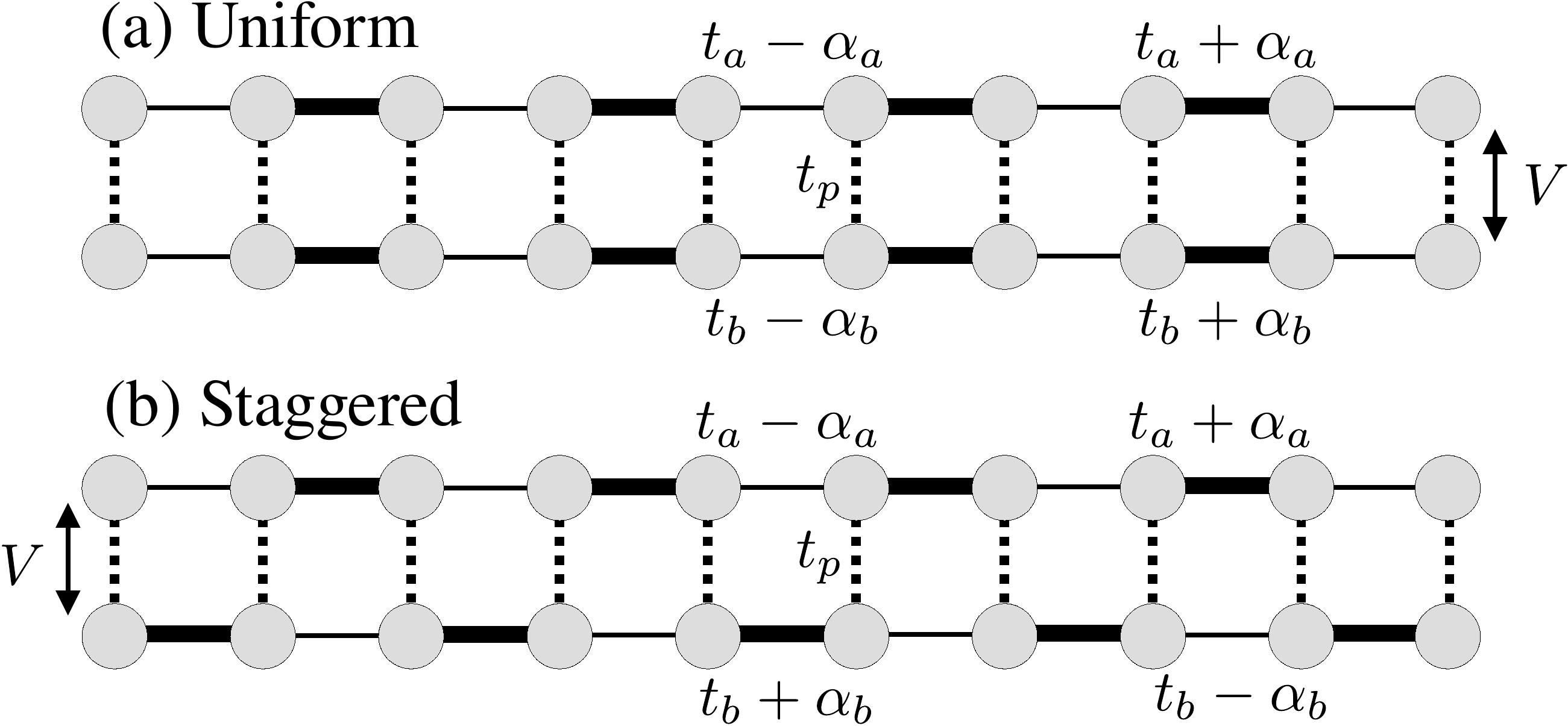}
 \end{center}
\caption{Two-leg ladder with different hopping dimerization patterns: (a) the uniform dimerization pattern, and (b) the staggered dimerization pattern. $t_a$, $t_b$ and $t_p$ are the hoppings along the leg-a, leg-b and along the rungs of the ladder. $\alpha_a$ and $\alpha_b$ decide the dimerization along leg-a and leg-b respectively. The circles represent the sites, and the thick (thin) bonds along the legs represent the strong (weak) hopping strengths. The dashed vertical lines illustrate the rung hopping. We allow inter-particle interaction $V$ only along the rungs which is marked by the arrow.}
\label{fig:lattice}
\end{figure}
The systems of interacting hardcore bosons on a two-leg SSH ladder studied here are described by the Hamiltonian
\begin{align}
\mathcal{H} = &-\sum_{j} (t_{a}-(-1)^{j}\alpha_a)  (a_{j}^{\dagger}a_{j+1}^{\phantom \dagger} + \text{H.c.})\nonumber\\
&-\sum_{j} (t_{b}-(-1)^{j}\alpha_b)  (b_{j}^{\dagger}b_{j+1}^{\phantom \dagger} + \text{H.c.})\nonumber\\
&- t_p\sum_{j}(a_{j}^{\dagger}b_{j}^{\phantom \dagger}+\text{H.c.}) + V\sum_{j} n_{aj}n_{bj}
\label{eq:ham}
\end{align}
where $a_j(b_j)$ and $n_{aj}$($n_{b_j}$) are respectively the bosonic annihilation and onsite number operator for $j^{\rm{th}}$ lattice site on leg-a(b). The hopping amplitudes along the respective legs are denoted by $t_a$ and $t_b$. The parameters $\alpha_a$ and $\alpha_b$ set the SSH model type dimerization along the $a$ and $b$ legs respectively. The two legs interact with each other through hopping and interactions along the rung-direction. Here $t_p$ represents the hopping amplitude along the rung, and $V$ is the inter-leg interaction strength. The hardcore constraint is imposed by assuming $(a^\dagger)^2=0$,  which ensures not more than one boson in a particular site.

Note that in the absence of both interaction ($V=0$) and rung hopping ($t_p=0$), the model considered above transforms to two decoupled legs whose physics solely depends on the choice of dimerization, i.e., $\alpha_a$ and $\alpha_b$. At half filling of hardcore bosons, the two chains exhibit topological or trivial character, when the dimerizations $\alpha_a$ and $\alpha_b$ are considered to be both positive or both negative. This indicates a topological phase transition as the dimerizations vary from negative to positive values or vice versa. Additionally, in the absence of dimerization, i.e. $\alpha_a=\alpha_b=0$, the non-interacting version ($V=0$) of $\mathcal{H}$ is known to exhibit a rung-Mott insulator (RMI) phase for any finite values of rung hopping $t_p$ where a particle tries to localize in each rung~\cite{crepin}. However, the combined effect of dimerization, rung hopping and interaction in such many-body system is not well explored.

In this work, we systematically analyze the effect of $t_p$ and $V$ on the topological properties of the system. In our studies, we consider two different combinations of dimerizations such as (a) uniform dimerization i.e. $ \alpha_a = \alpha_b > 0$ and (b) staggered dimerization i.e. $ \alpha_a = -\alpha_b > 0$ as depicted in Fig.~\ref{fig:lattice}(a) and (b), respectively. In the absence of $t_p$ and $V$, for the uniform dimerization case, the particles in both the legs are in  topological phases and for the staggered dimerization case the particles in leg-a are in topological phase and in leg-b they are in the trivial phase. In the following we study the effect of $t_p$ and $V$ on the topological features of the system.


To achieve the above two dimerization patterns, we set the following parameters. The uniform dimerization is enforced by fixing $t_a = t_b = t$ and $\alpha_a = \alpha_b = \alpha$. Similarly, the staggered dimerization pattern along the legs is enforced by setting $t_a = t_b = t$ and $\alpha_a = -\alpha_b = \alpha$. We consider $t=0.6$ and $\alpha = 0.4$ in our calculations such that the hopping strength in strong and weak bonds in the legs are $t_1 = t+\alpha = 1$ and $t_2 = t-\alpha = 0.2$, respectively. Here $t_1=1.0$ defines the unit of energy, which makes the other parameters in the system dimensionless. These choices of parameters make the system strongly dimerized in the decoupled limit ($t_p=V=0$) where the topological properties are very prominent such as the edge states with very small correlation length (well localized) and large bulk gap. 

We numerically obtain the ground state properties of the system using the density matrix renormalization group (DMRG) method~\cite{whitedmrg, ulrich2004} based on the matrix product states (MPS) approach~\cite{rommer, SCHOLLWOCK201196}. We consider a ladder of length up to $L=100$ rungs (i.e. $200$ sites) for different particle number sectors of $N$ hardcore bosons with the desired densities $\rho=N/2L$. Unless otherwise mentioned, we compute the physical quantities in the thermodynamic limit ($L=\infty$) by using the appropriate extrapolation technique to minimize finite-size effects. The accuracy of the DMRG simulations is ensured by considering sufficiently large bond dimensions ($\chi$) up to $\chi=500$.

\section{Single Particle Spectrum}
\label{sec:single}
Before presenting many-body features, we first analyze the single particle spectrum of the model shown in Eq.~\ref{eq:ham}. Although the model under consideration is for hardcore bosons, we show that significant insights about the topological properties of the many-particle ground state can be obtained from the single particle spectrum at different fillings. In the following, we describe the quantum phases at different fillings for both the uniform and staggered dimerization configurations considered in our study.
\label{sec:re}
\begin{figure}[t]
\includegraphics[width=1\columnwidth]{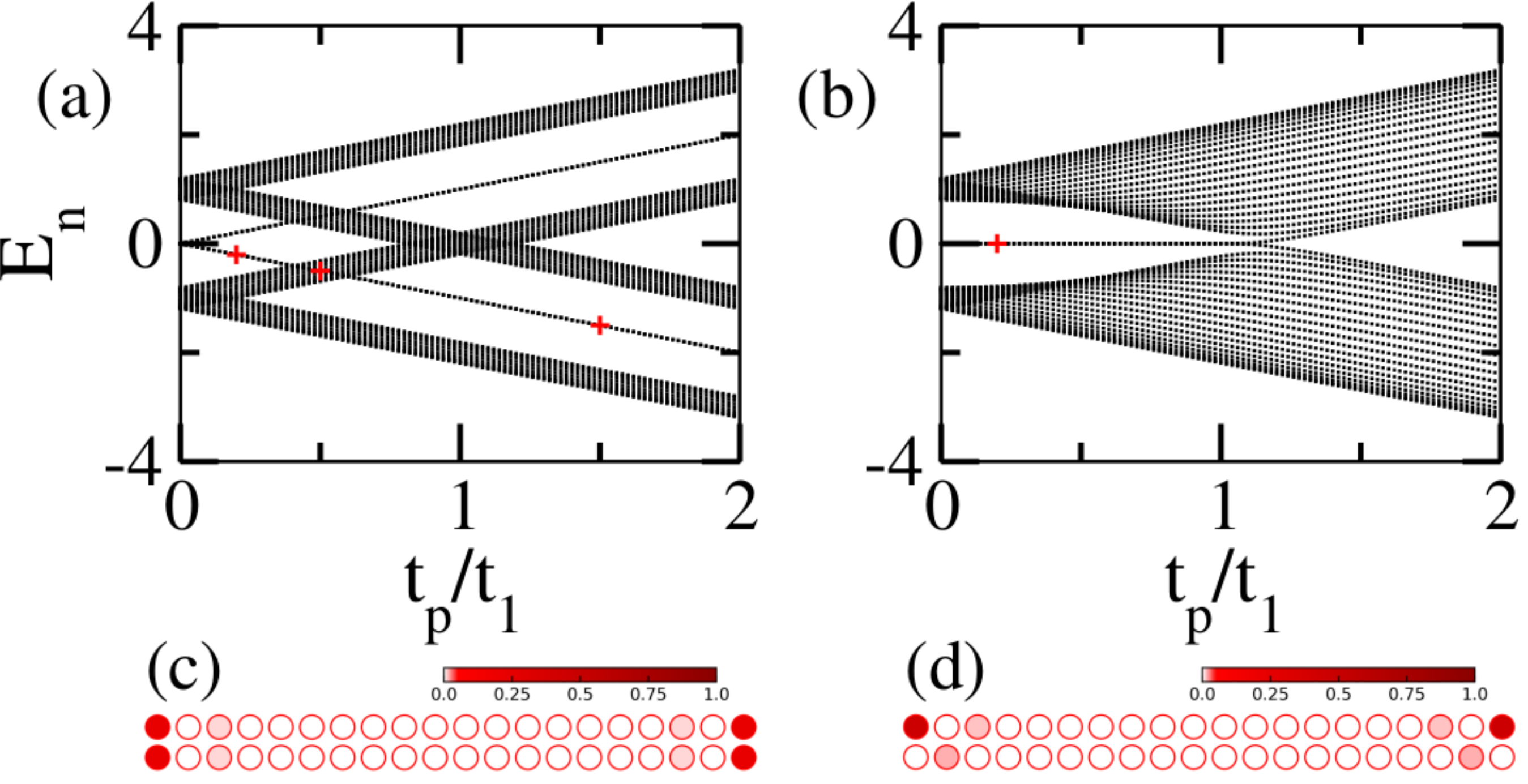}
\caption{The single particle energy spectrum of a system of $40$ sites($L=20$) with varying $t_p$ for (a) uniform dimerization pattern ($t_a = t_b = 0.6t_1$ and $\alpha_a=\alpha_b = 0.4t_1$) and (b) staggered dimerization pattern ($t_a = t_b = 0.6t_1$ and $\alpha_a = -\alpha_b = 0.4t_1$) along the legs. (c) and (d) show the on-site probability $|\psi_j|^2$ of states marked by the red plus signs in (a) and (b) respectively. Note that, the on-site probability for of all the states marked in (a) are identical as that of in (c). This clearly indicates that the edge states are crossing the band from one gapped phase to another.}
\label{fig:single}
\end{figure}

As a specific representative case, we plot the numerically obtained single particle spectrum as a function of $t_p$ for a system of length $L=20$ for uniform 
dimerization($t_a = t_b = 0.6t_1$ and $\alpha_a=\alpha_b = 0.4t_1$) and staggered
dimerization case ($t_a = t_b = 0.6t_1$ and $\alpha_a = -\alpha_b = 0.4t_1$) in Fig. \ref{fig:single}(a) and (b) respectively.
Note that for the case of uniform dimerization, the system is a weak topological system consisting of a stack of lattices that are topologically identical in nature. When $t_p = 0$, the two legs of the ladder are equivalent to two isolated SSH chains, which are individually known to manifest a gap in the middle of the spectrum due to the onset of the bond order (BO) phase~\cite{Mondal2019}. This gapped phase hosts a pair of symmetry protected zero-energy edge states (see Fig.~\ref{fig:single}(a)). When $t_p$ takes finite values, apart from the gap at the middle of the spectrum, two more gaps open up symmetrically at quarter and three quarter fillings of the spectrum after a critical $t_p$ due to the formation of a plaquette order (PO) in the system, which will be discussed in details in the next section.
The zero energy edge modes 
in the two legs of the ladder hybridize to become energetic while staying at the edges. These states can be called as mid-gap states. However, in this case, these states merge with the energy bands before the bulk gap closes at a higher value of $t_p$ violating the bulk-edge correspondence. As mentioned before, this phenomenon can be attributed to the  weak topology of the uniform dimerization configuration. With further increase in $t_p$, the bulk gap opens up again but without the mid-gap states. A pair of mid-gap states appear \sout{again} inside the two symmetrically formed gaps at quarter and three quarter fillings of the energy spectrum.
To confirm that these mid-gap states are edge states, in Fig~\ref{fig:single}(c) 
we plot the probability density $|\psi_j|^2$ of these  states marked by red plus signs in Fig.~\ref{fig:single}(a) where $\psi_j$ is the wavefunction amplitude at each lattice site $j$. It can be clearly seen that the  amplitude $\psi_j$ has maximum weight on the edge lattice sites of both the legs. 
We note that these mid-gap states cross the bulk bands and go from one gapped phase to another gapped phase as a function of $t_p$. While crossing the band, the edge state retains its properties, i.e., the probability density behaves similar to Fig.~\ref{fig:single}(c) at all three positions marked by the red plus symbols in Fig.~\ref{fig:single}(a).

For the case of staggered dimerization, however, the single particle spectrum exhibits a simple but topologically richer picture. In this case a gap-closing transition occurs between two gapped phases as a function of $t_p$ where one of the gapped phases hosts zero energy edge modes as can be seen from Fig.~\ref{fig:single}(b). This indicates a clear topological phase transition where the bulk-boundary correspondence is preserved. We plot the probability density ($|\psi_j|^2$) in Fig.~\ref{fig:single}(d) for the states marked by the red plus sign in 
Fig.~\ref{fig:single}(b). It can be seen that the edge states are concentrated on the two edges of leg-a.

The single-particle orbitals discussed above can be filled one by one from the lowest energy onwards to obtain different ground state phases arising at different fillings of fermions in the many-body context. Fundamentally, hardcore bosons share some properties with the spinless fermions such as the energy and diagonal correlations but strictly in one dimension. The situation is different in the case of a two-leg ladder. A critical study on this front has been performed  in Ref.~\cite{crepin} where the differences between non-interacting spinless fermions and hardcore bosons in a two-leg ladder model with uniform leg hopping has been highlighted. It has been shown in Ref.~\cite{crepin} that at half filling, both spinless fermions and hardcore bosons stabilize to the gapped phases as a function of the rung hopping. In these gapped phases the particles prefer to localize on the rungs. However, the difference between fermions and hardcore bosons is that the transition to the gapped phase occurs after a critical $t_p$ for the case of spinless fermions, whereas for hardcore bosons, the system becomes a gapped RMI phase for any finite value of $t_p$ indicating a clear difference between the two systems. While these differences exist, the single-particle analysis provides a broad basis for understanding topological insulators in both fermionic and bosonic systems.

In our studies, we show that for the model under consideration, the hardcore bosons in the many-body limit exhibits different phases, such as, topological bond order (BO) phases, rung-Mott insulators (RMI), plaquette order (PO) phases along with superfluid (SF) phases at incommensurate fillings. We first discuss the scenario without interaction ($V=0$) and then with interaction ($V\neq 0$) for both the cases of uniform and staggered dimerizations.

\section{Many-body phases }
\label{sec:many}
Interacting bosons in periodic lattices exhibit a range of rich phenomena~\cite{bloch_rev}. One such phenomenon is the phase transition from the gapless superfluid (SF) to gapped Mott insulator (MI) phase at integer filling~\cite{Fisher89, jaksch,bloch}. While the SF-MI transition has been extensively studied in all the three dimensions, in one dimensional lattices, the SF phase is characterised by off-diagonal quasi-long-range order. In the presence of strong onsite interaction, i.e., in the hardcore limit, the system reaches the Tonk's limit~\cite{Bloch_paredes_expt}. Such hardcore bosons in one dimension mimic the physics of spin polarized fermions and therefore, the physics of the associated many-body problem can be extracted from the single particle picture. One of the simplest examples is the one-dimensional bosonic SSH model, which exhibits a topological phase transition in presence of interaction~\cite{fleischhauer_2013_prl,Mondal2019,Mondal2020}. However, in the case of quasi-1D lattices, this mapping is not straightforward and therefore it is difficult to predict the physical picture of a many-body problem. 

In this section we analyse the many-body physics of the SSH ladder shown in Eq.~\ref{eq:ham} in the absence of interaction (i.e. $V=0$) for both the cases of dimerizations considered.  
 
\subsection{Uniform  dimerization}\label{sec:re_idV0}
\begin{figure}[t]
\includegraphics[width=1\columnwidth]{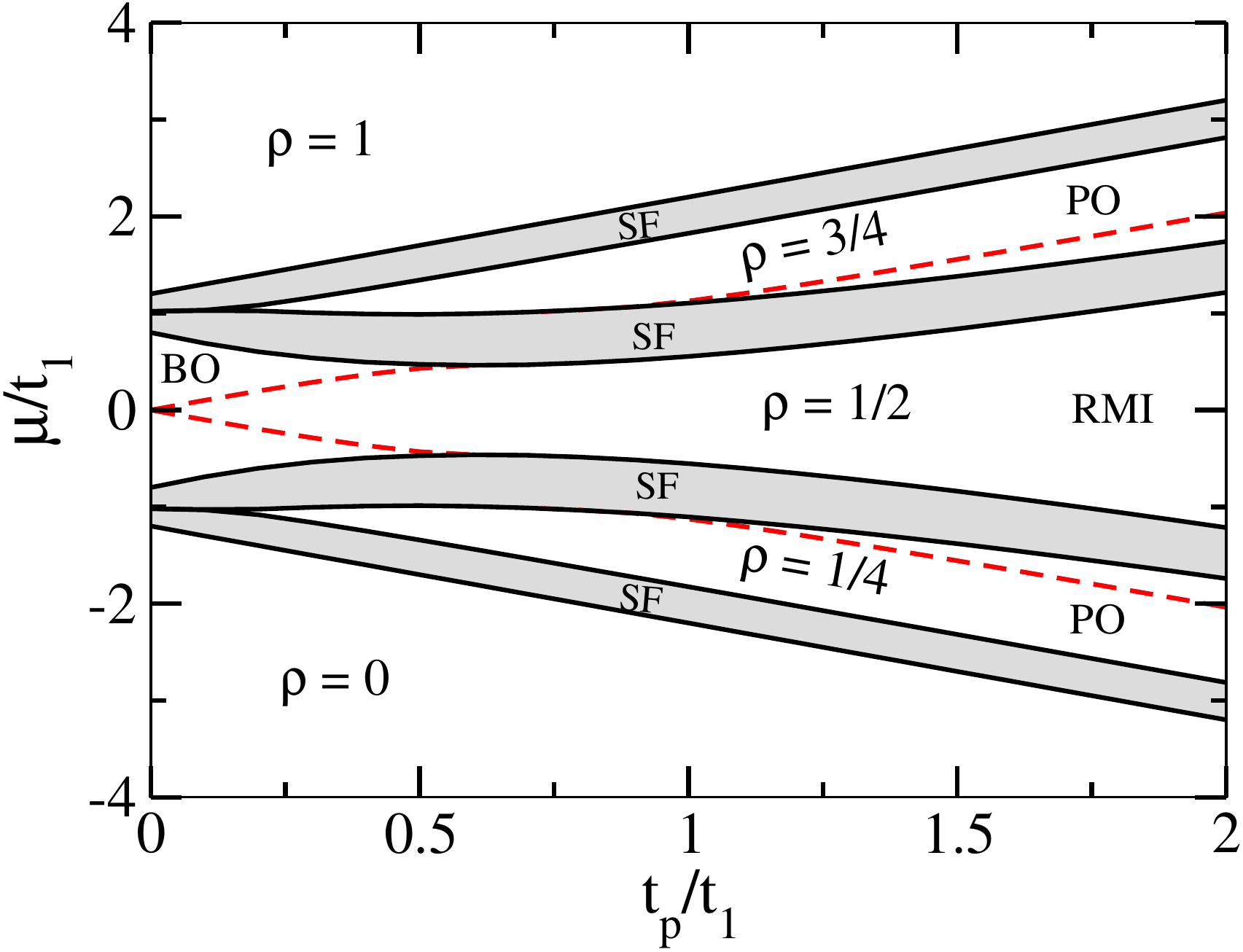}
 \caption{Phase diagram of Model(\ref{eq:ham}) with $V=0$, $t_a = t_b = 0.6t_1$ and 
 $\alpha_a=\alpha_b = 0.4t_1$ plotted in the $t_p$-$\mu$ plane. The solid black lines show the 
 phase boundaries of gapped phases, such as the plaquette order (PO), the bond order (BO), and the rung-Mott insulator (RMI) phases. The dashed red lines denote the mid-gap states. The gapless superfluid (SF) phase is represented by the grey shaded area.}
 \label{fig:samePD}
\end{figure}

\begin{figure}[t]
\includegraphics[width=1\columnwidth]{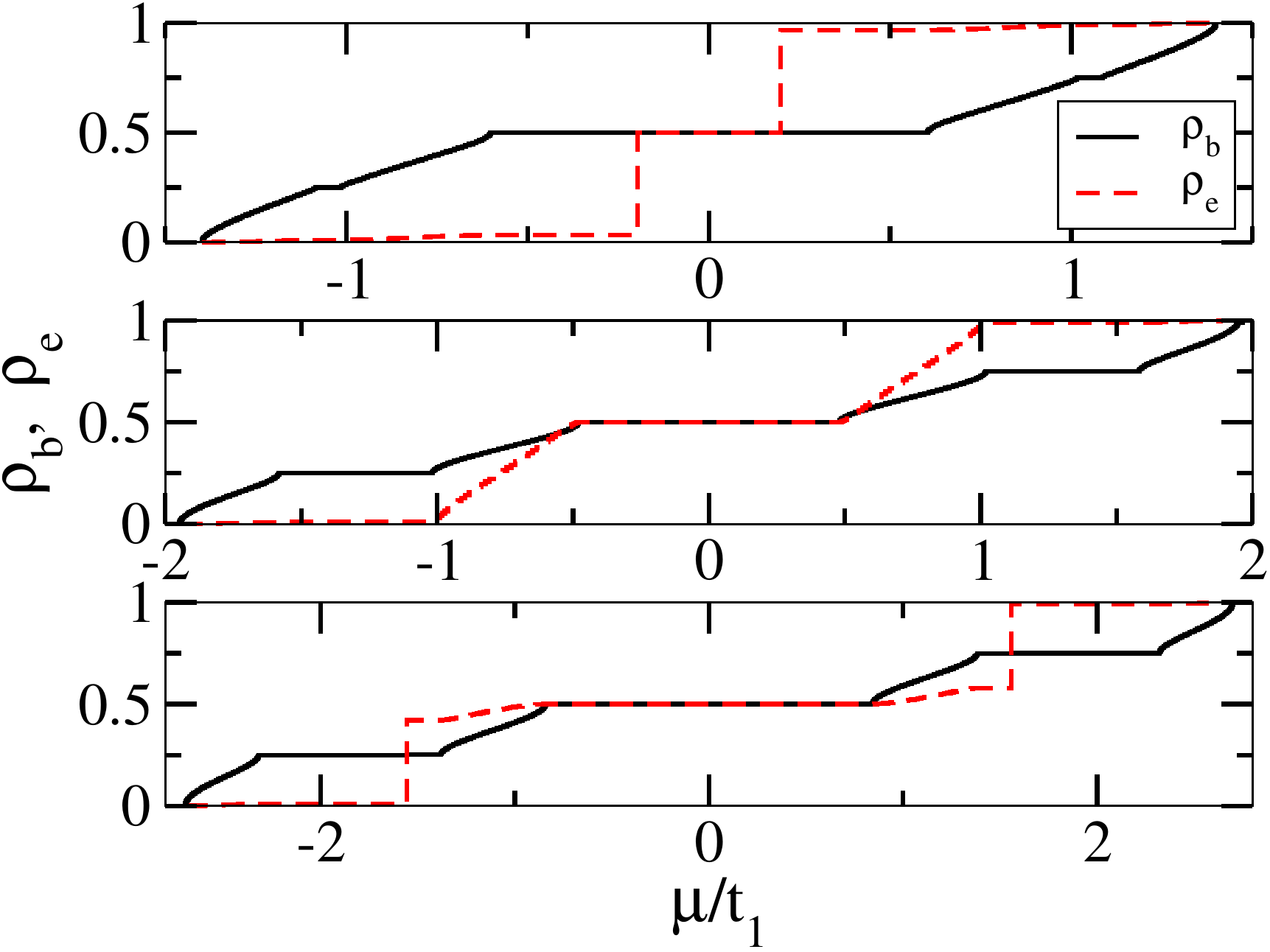}
\caption{The figure displays the bulk and edge densities of the system represented by $\rho_b$ and $\rho_e$ respectively as a function of $\mu$. (a), (b) and (c) show the $\rho_b$ (solid lines) and $\rho_e$ (dashed lines) with varying chemical potential $\mu$ for $t_p=0.2t_1,~0.75t_1$ and $1.5t_1$, respectively, corresponding to three cuts in the phase diagram shown in Fig~\ref{fig:samePD}. This shows the nature of the bulk phases (gapped or gapless) when the edge states are being filled in different parameter regimes.}
\label{fig:rhomu}
\end{figure}

In Fig.~\ref{fig:samePD} we show the phase diagram for uniform dimerization on both the legs of the ladder with $t_a = t_b = 0.6t_1$ and $\alpha_a=\alpha_b = 0.4t_1$. The phase diagram is obtained as a function of the chemical potential $\mu$ and $t_p$, and it exhibits three gapped phases at $\rho=1/2,~1/4$ and $3/4$ denoted by the white regions with black solid boundaries. The boundaries of these gapped phases  are calculated using the formula
\begin{equation}
 \mu_- = E(N)- E(N-1);~~ \mu_+ = E(N+1)- E(N),
\end{equation}
where $E(N)$ is the ground state energy of the system consisting of $N$ bosons. Here, $\mu_-$ ($\mu_+$) is the chemical potential that defines the lower (upper) boundary of a gapped phase. All the lines in the figure are plotted after extrapolating the values of $\mu_{\pm}$ to thermodynamic limit.
One of the major differences that appears in the energy gap at $\rho=1/2$ as compared to the single particle picture is that it always remains finite as a function of $t_p$ (compare with Fig~\ref{fig:single}(a)). A similar feature exists for the non-dimerized hardcore bosonic ladder where a gapped RMI phase extends until $t_p=0$  at half filling in contrast to the single-particle physics~\cite{crepin}.
The red dashed lines are the mid-gap edge states which are determined by analysing the $\rho$-$\mu$ plot, as shown in 
Fig.~\ref{fig:rhomu}. The figure depicts the behaviour of the bulk ($\rho_b$) and edge ($\rho_e$) densities   which are defined as 
\begin{equation}
\rho_b = \frac{1}{L-2}\sum_{j=2}^{L-1}\langle n_{aj} + n_{bj}\rangle
\end{equation} 
and 
\begin{equation}
\rho_e = \frac{1}{4}\langle n_{a1} + n_{b1} + n_{aL} + n_{bL}\rangle
\end{equation}
respectively as a function of $\mu$. The panels (a), (b) and (c) of Fig.~\ref{fig:rhomu} show the $\rho$-$\mu$ behaviour for $t_p= 0.2t_1,~0.75t_1$ and $1.5t_1$ respectively corresponding to three different cuts in the 
phase diagram of Fig.~\ref{fig:samePD}. 
The plateaus in the $\rho_b$ curves (black solid lines) indicate the three bulk gaps at $\rho=1/4$, $\rho=1/2$ and $\rho=3/4$ which confirm the existence of the gapped phases. The shoulders around the plateaus are the signature of gapless regions where the system exhibits a superfluid (SF) character with off-diagonal quasi-long-range order.

The behaviour of the $\rho_e$ (red dashed lines) in Fig.~\ref{fig:rhomu} reflects the existence of edge states in the system. When the bulk is gapped, the sudden change in the number of particles at the edges ($N_e$) by two or a change in $\rho_e$ by $0.5$ for $t_p=0.2t_1$ and $1.5t_1$ (Fig.~\ref{fig:rhomu} (a) and (c)) mark the $\mu$ of the mid-gap edge states. The change in $N_e$ by two particles is due to the two degenerate mid-gap edge states at that particular $\mu$.
Interestingly, for the case of $t_p = 0.75t_1$ (Fig.~\ref{fig:rhomu}(b)),  $\rho_e$ increases continuously only when the bulk is in the gapless SF region with density $1/4 < \rho_b < 1/2$ and $1/2 < \rho_b < 3/4$. 
This indicates that, the edge states get filled up while the bulk of the system exhibits a gapless SF phase. As discussed earlier, a similar phenomenon can be seen in the single particle case, where the edge states survive inside the bands between half and quarter fillings.

\begin{figure}[t]
\includegraphics[width=1\columnwidth]{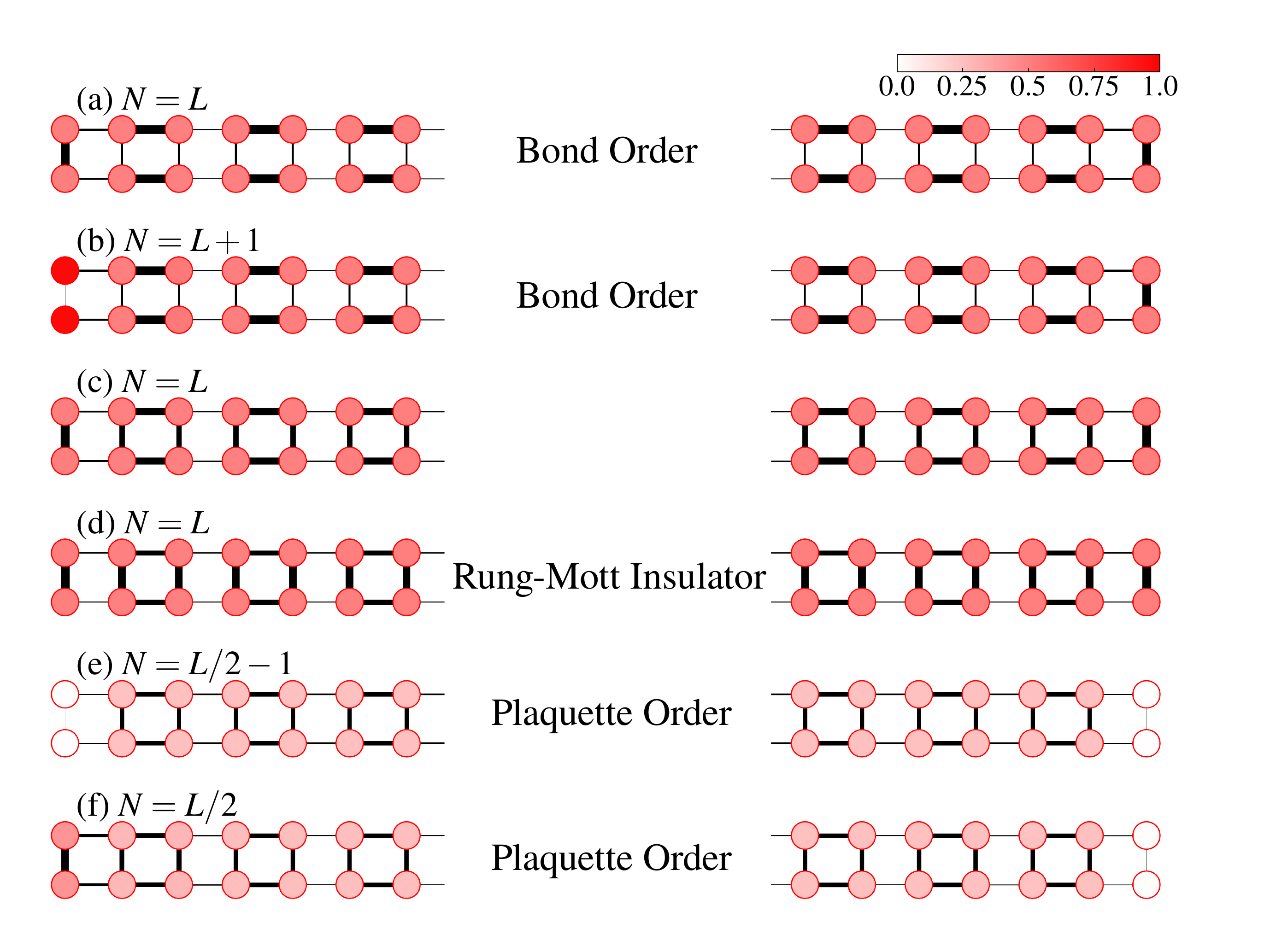}
 \caption{The figure shows the bond energies ($B_j$) of all the bonds defined in Eq.~\ref{eq:bondener} and the onsite particle number ($\langle n_j\rangle$) for different phases corresponding to Fig.~\ref{fig:samePD} with a system consisting of $40$ sites ($L=20$ rungs). In the figures, the thickness of a bond is proportional to the respective strength of $B_j$, and the face colour of the circles represents the values of $\langle n_j\rangle$. This captures different dimerization patterns and the existence of edge states in different phases. (a) shows $B_j$ and $\langle n_j\rangle$ corresponding to the bond order (BO) phase at $1/2$-filling ($N=L$) for $t_p=0.2t_1$, which has two filled edge states (localized at each edge). The parameters in (b) are the same as (a) but with $N=L+1$, which has three occupied edge states (two localized on the left edge and one on the right edge). (c) and (d) show the same quantities for $t_p=0.2t_1,~ 0.75t_1$ and $1.5t_1$ respectively at $1/2$-filling ($N=L$). The change in bond order pattern can be seen going from the bond order (BO) (a) to the rung-Mott insulator (RMI) (d) phase. (e) and (f) correspond to the $1/4$-filling plaquette order (PO) phases for parameters $t_p=0.5t_1$ with $N=L/2-1$ and $t_p=1.5t_1$ with $N=L/2$, respectively. The edge state appears (localized on the left edge) in (f). Note that in all the cases, we have used a small onsite potential of $-0.001t_1$ in one edge to break the degeneracy of the 
 edge-state pair.}
 \label{fig:BO}
\end{figure}

Now we discuss the properties of the gapped phases in different parameter regimes. At $\rho=1/2$, when $t_p = 0$, the two chains are isolated SSH chains. Due to the dimerized hopping on both legs, the system exhibits a gapped bond order (BO) phase along the legs with edge states on both legs. For the other limit of $t_p$, when it is large compared to the hopping along the legs the system is gapped once again, but due to the dimerization along the rungs. This phase is known as the rung-Mott insulator (RMI) phase as already introduced before. However, from the phase diagram shown in Fig.~\ref{fig:samePD}, it can be seen that the two gapped phases at two limits of $t_p$ are connected without any gap closing. To identify these phases, we calculate the bond energy along the legs and the rungs using the formula given by
\begin{align}
B_{a,j} = & \langle a_j^\dagger a_{j+1} + H.c.\rangle ;~ B_{b,j} = \langle b_j^\dagger b_{j+1} + H.c.\rangle\nonumber\\
B_{r,j} = & \langle a_j^\dagger b_j + H.c.\rangle
\label{eq:bondener}
\end{align}
Here the subscripts $a$, $b$ and $r$ denote the leg-a, leg-b and the rung respectively. Fig.~\ref{fig:BO}(a-d) display the bond energy in all the bonds along with the onsite particle densities for a system 
of size $L=20$ at half filling when $t_p=0.2t_1,0.2t_1,0.75t_1$ and 
$1.5t_1$ respectively. Here the thickness of the bonds is proportional to the bond energy 
$B_j$, and the face colour of the circles represents the particle densities on a particular site. In Fig.~\ref{fig:BO}(a) we can see that the alternate bonds along the legs are stronger indicating the BO phase with one particle delocalized in the first and last rungs. When both the legs are isolated($t_p=0$), a configuration with one particle in one of the edges of leg-a and one particle in one of the edges of leg-b, is an energetically favourable configuration at half filling. However, in Fig.~\ref{fig:BO}(a) since we consider a finite rung hopping($t_p=0.2t_1$), the system chooses to be in a state such that both the edge rungs have one particle delocalized in them, which is the minimum energy configuration. We can get more insights regarding the edge states by moving away from half filling, e.g., by adding one extra particle to the system we get a strong bond on the right most rung as shown in Fig.~\ref{fig:BO}(b). This is because at $\rho_b=1/2$, there are in principle two pairs of edge states at different $\mu$. Since in this case we consider $N=L+1$ particles, three out of four edge states are filled. Due to this, two particles reside on the left edge of the ladder (i.e. the first rung) and 
one particle resides on the right edge of the ladder (i.e. the last rung). The particle that resides on the last rung delocalizes itself on the rung due to finite $t_p$ (as can be seen from Fig.~\ref{fig:BO}(b)).

Apart from the gapped phase at $\rho=1/2$, we also see two more gapped phases at $\rho=1/4$ and $3/4$, which is expected from the single particle analysis. Note that the gapped phases at $\rho=1/4$ and $\rho=3/4$ are dual to each other due to the particle-hole symmetric nature of the model in the absence of interaction ($V=0$). At $\rho=1/4$ ($\rho=3/4$), for $t_p=0$, the system is in SF phase but when $t_p$ is strong, a particle (hole) tends to get localized in every alternate plaquette of the ladder and hops within the plaquette only. This phase is very similar to the BO phase but here a particle gets trapped in a plaquette rather than a bond, which we call the plaquette order (PO) phase. In Fig.~\ref{fig:BO}(e) and (f), we show the bond energy and particle densities of the gapped phase at $\rho=1/4$ for $t_p = 0.5t_1$ and $1.5t_1$ respectively. Here we can see that at every alternate plaquette, the bond energy is stronger. But the edge state exists only in 
Fig.~\ref{fig:BO}(f), which is dimerized in the first rung.

This analysis clearly shows that the strong topological character ceases to appear in the case of uniform dimerization and the zero energy edge states become mid-gap edge states for any finite value of $t_p$, hence ruling out any topological phase transition as a function of $t_p$. However, in the following we show that the case of staggered dimerization exhibits a clear topological phase transition even in presence of finite interaction.

\subsection{Staggered dimerization}\label{sec:re_op}
In this case, the leg-a (leg-b) of the ladder is considered to be of topological (trivial) nature.  Similar to the uniform dimerization case, we first discuss the many-body phase diagram in the absence of interaction ($V=0$) and compare it with the single particle picture (Fig.~\ref{fig:single}(b)).
\begin{figure}[t]
\includegraphics[width=1\columnwidth]{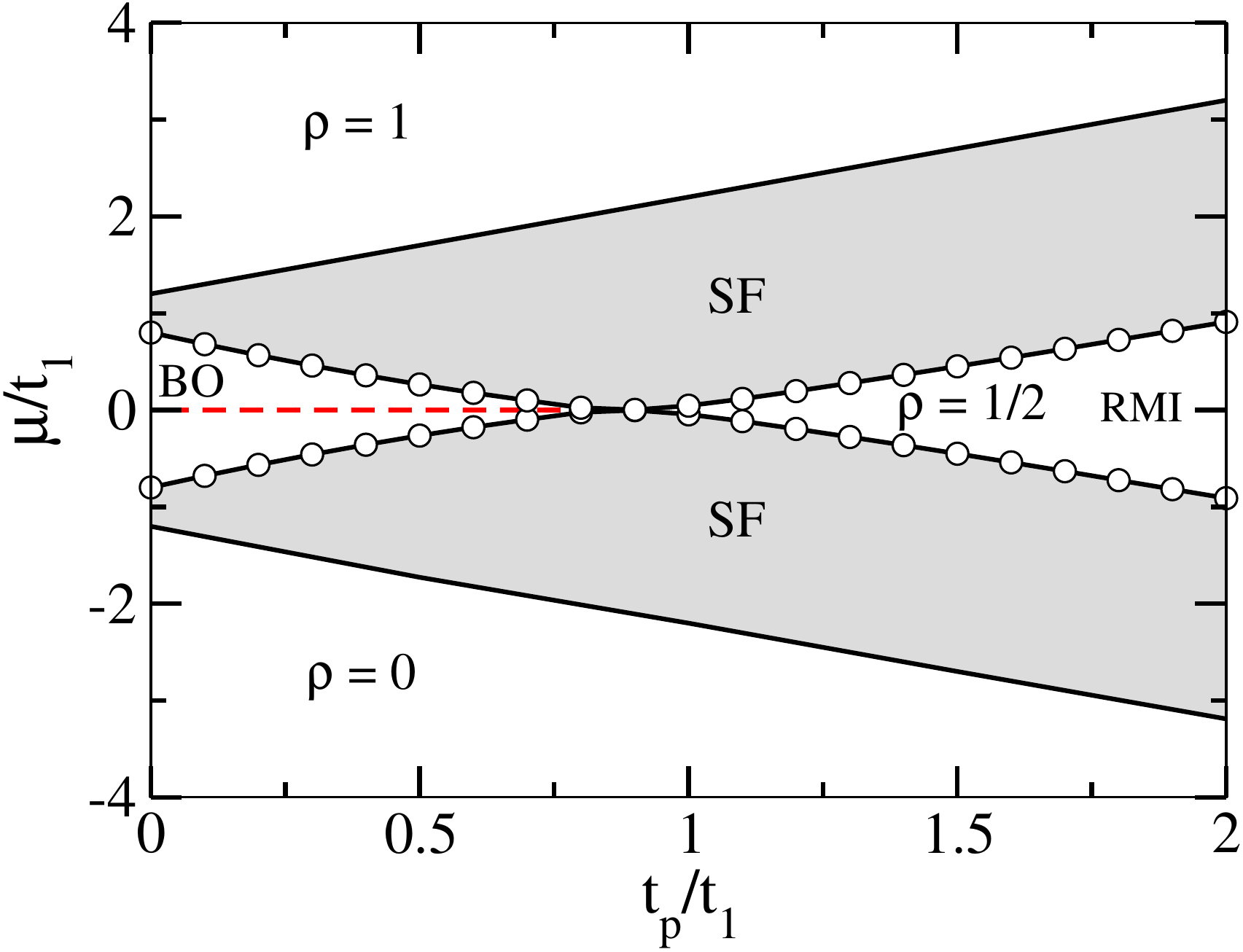}
 \caption{Phase diagram of Model(\ref{eq:ham}) with $V=0$, $t_a = t_b = 0.6t_1$ and 
 $\alpha_a=-\alpha_b = 0.4t_1$ is plotted in $t_p$-$\mu$ plane. The solid black lines show the 
 phase boundaries of gapped phases and the dashed lines denote the mid-gap states. The gapless superfluid (SF) phase is represented by the shaded grey area. A topological phase transition happens through a gap-closing point for $\rho=1/2$ from a non-trivial bond order (BO) to a trivial rung-Mott insulator (RMI) phase.}
 \label{fig:oppPD}
\end{figure}
In Fig.~\ref{fig:oppPD}, we show the phase diagram with the parameters $t_a = t_b = 0.6t_1$ and $\alpha_a = -\alpha_b = 0.4t_1$ in the $\mu-t_p$ plane. In the figure, the white regions bounded by the black line with circles correspond to the gapped phases, and the grey regions around them are the gapless regions. The phase diagram clearly shows a phase transition between two gapped phases as a function of $t_p$ at half-filling. In the following, we show that this transition through a gap-closing point is well defined topological phase transition, which was already indicated in the single-particle picture (compare with Fig.~\ref{fig:single}(b)). 

\begin{figure}
\includegraphics[width=1\columnwidth]{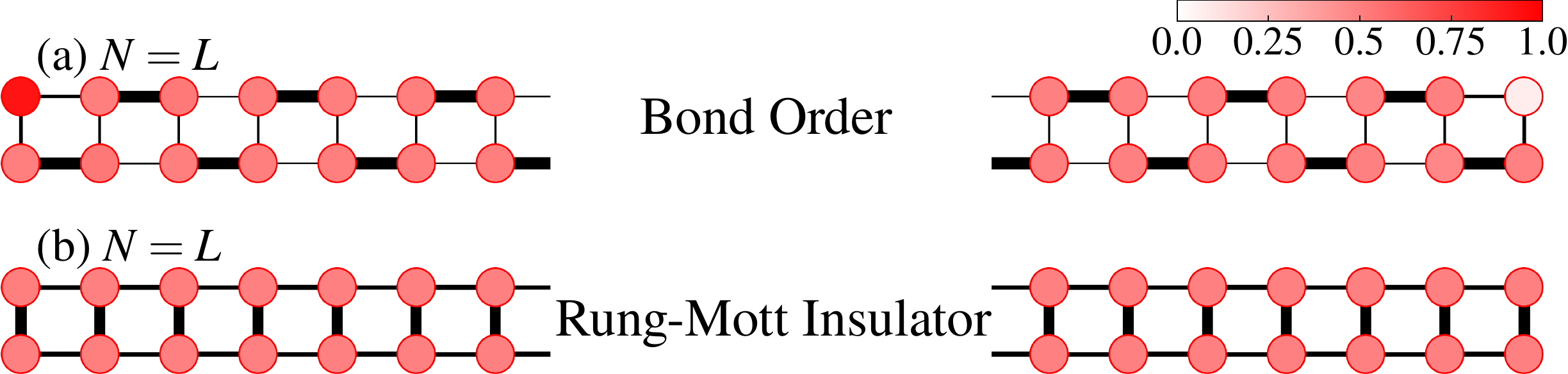}
 \caption{The figure shows the bond energy ($B_j$) of all the bonds defined in Eq.~\ref{eq:bondener} and the onsite particle number ($\langle n_j\rangle$)  for different phases of a system consisting of $40$ sites ($L=20$). (a) and (b) represent two gapped phases at the $1/2$-filling ($N=L$), which are topological bond order (BO) and trivial rung-Mott insulator (RMI) phases corresponding to the parameter value $t_p=0.25t_1$ and $1.5t_1$ respectively in Fig.~\ref{fig:oppPD}. Here the thickness of a bond is proportional to the corresponding value of $B_j$, and the face color of the circles represents the onsite particle number. We can see the change in dimerization pattern between (a), which also has an occupied edge state (localized at the left edge), and (b).  Note that in both cases, we have used a small onsite potential of $-0.001t_1$ in one edge to break the degeneracy of the edge-state pair.}
 \label{fig:BO2}
\end{figure}

First of all, when $t_p = 0$, the system corresponds to two isolated SSH chains. Due to staggered dimerization, the upper leg at $\rho=1/2$, exhibits a nontrivial BO phase with zero-energy 
edge states,  and the lower leg exhibits a trivial BO phase without any edge states. Because of this, there exists a finite gap at $t_p=0$ with zero energy states at the middle. Upon switching on the rung hopping $t_p$, the system shows an affinity towards dimerizing along the rungs due to the enforcement of the RMI character. As a result, the topological BO phase and the edge states (red dashed line) survive up to certain values of $t_p$. After a critical value of $t_p \sim 0.9t_1$, the dimerization in the rungs dominate, leading to the appearance of the gapped RMI phase, which does not exhibit any edge states. This transition clearly occurs through a gap closing point, indicating a topological phase transition as a function of $t_p$.
These two gapped phases can be distinguished by comparing their bulk behavior through $B_j$ as shown in Fig.~\ref{fig:BO2}(a) 
and (b) which are plotted for parameters $t_p=0.25t_1$ and $1.5t_1$ respectively at $\rho=1/2$. From the figure, it can be seen that when $t_p$ is below the critical value, 
the dimerization is maximum along the legs of the ladder and the left edge 
of leg-a is occupied by a particle, indicating the presence of an edge state. When $t_p$ is above the 
critical value, we see a dominant dimerization along the rungs forming the RMI phase, which does not host any edge states. 

At this point, it is evident that the staggered dimerization case favors a topological phase transition in the many-body limit, which is absent in the case of uniform dimerization. 
In the remaining part of the paper, we focus on the role of interaction on the topological character displayed by the many-body system in the staggered dimerization case.  

\

\section{Interaction induced topological phase transition}
\label{sec:inter}
\begin{figure}[t]
\includegraphics[width=1\columnwidth]{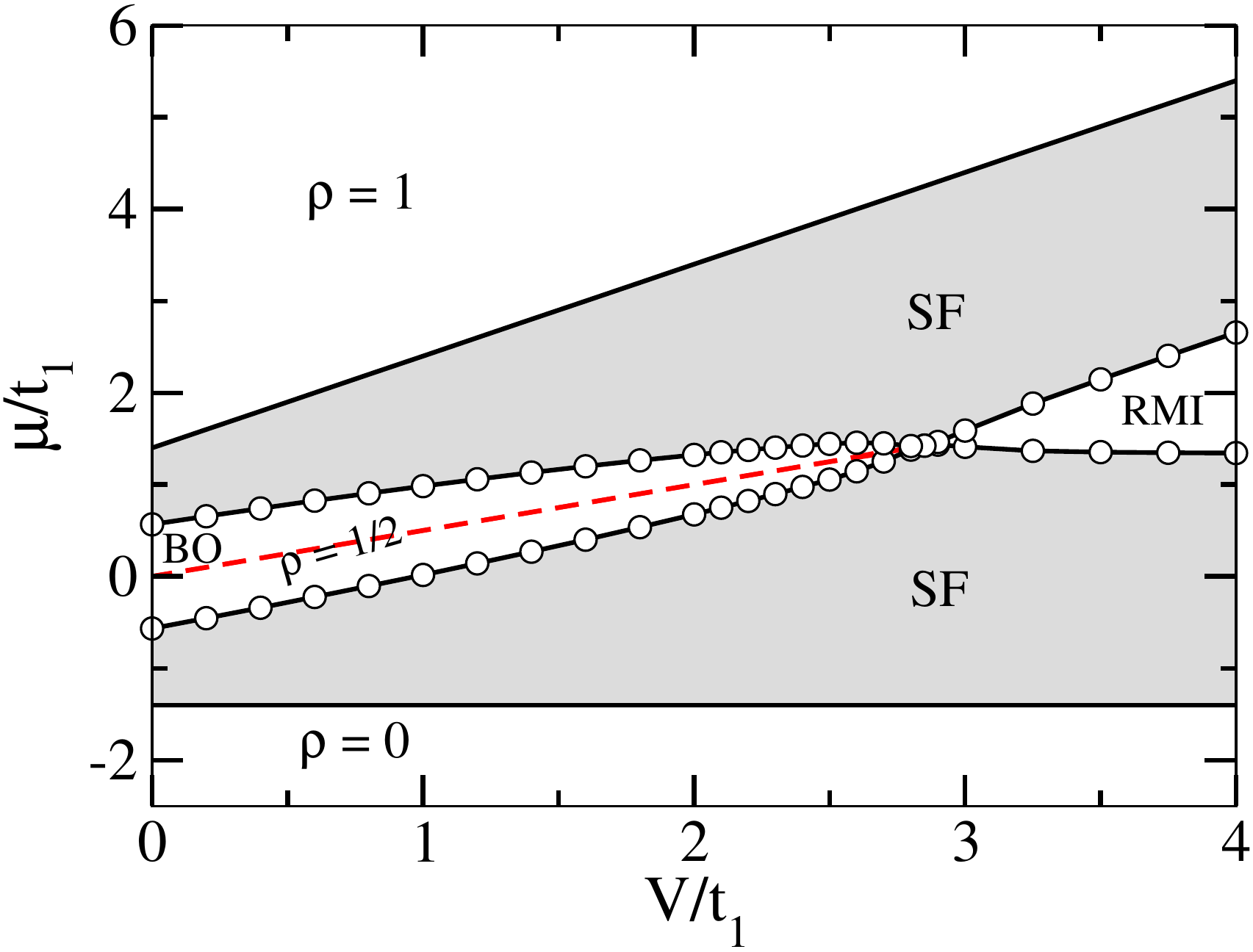}
 \caption{Phase diagram of Model(\ref{eq:ham}) with $t_a = t_b = 0.6t_1$, 
 $\alpha_a=-\alpha_b = 0.4t_1$ and $t_p=0.2t_1$ is plotted in $V$-$\mu$ plane. The solid black lines show the phase boundaries of gapped phases and the red dashed lines stand for the mid-gap edge states. The gapless superfluid (SF) phase is represented by the shaded grey area. Here, at $\rho=1/2$, the topological phase transition occurs  from a non-trivial bond order (BO) to a trivial rung-Mott insulator (RMI) phase through a gap-closing point with increasing $V$.}
 \label{fig:oppPDV}
\end{figure}

In this section, we study the fate of the topological phase that occurs for the staggered dimerization case in the presence of interaction ($V$).
For this purpose, we start in the parameter domain $t_p = 0.2t_1$, $t_a = t_b = 0.6t_1$ and $\alpha_a=-\alpha_b = 0.4t_1$ in the non-interacting phase diagram shown in Fig.~\ref{fig:oppPD}, where the system is in a gapped BO phase exhibiting edge states at $\rho=1/2$ and examine the effect of finite $V$. By introducing $V$, we obtain a phase diagram which is shown in Fig.~\ref{fig:oppPDV} in the $\mu-V$ plane, obtained after appropriate finite-size extrapolation. The phase diagram clearly shows a gapped to gapped phase transition where the gapped and gapless phases are indicated by the white and grey regions, respectively. We find that the degenerate edge states still survive at finite $V$ (indicated by the red dashed line), preserving the topological nature of the phase. With increase in $V$, however, a gap-closing transition to a trivial phase occurs  at a critical interaction strength $V_c \sim 2.85t_1$. The gap closing transition induced by interaction signifies the bulk-edge correspondence of the topological phase transition even though the edge states shift from the $\mu/t_1=0$ value due to the particle-hole symmetry breaking of the Hamiltonian. Note that in this case, the gapped phase below $V_c$ is a nontrivial BO phase, and above $V_c$, it is the RMI phase. 

\begin{figure}
\includegraphics[width=1\columnwidth]{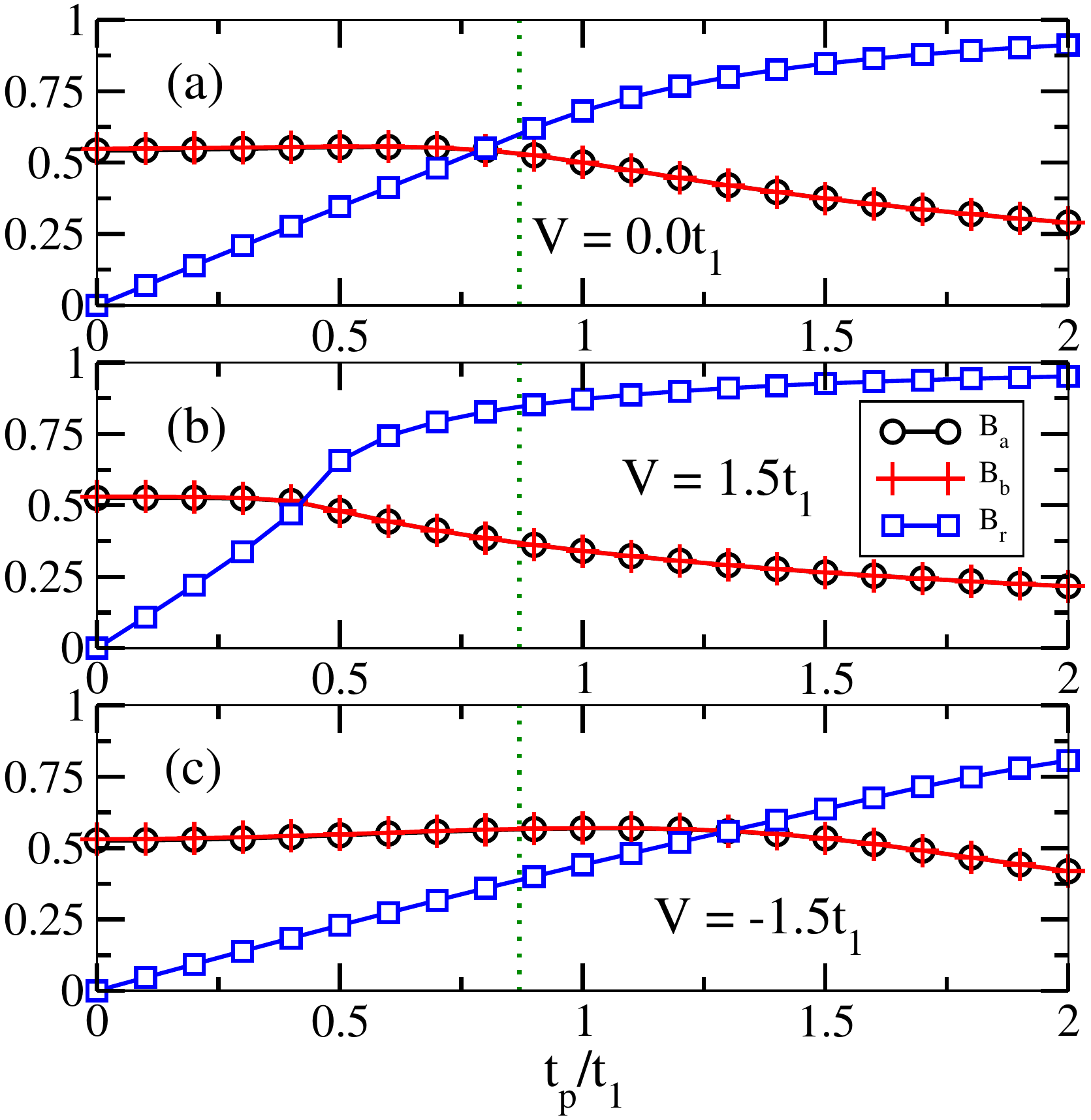}
 \caption{Bond energies along the legs ($B_{a,b}$) and along the rung ($B_r$) computed by averaging over all the respective bonds for 240 sites ($L=120$) at $\rho=1/2$. (a), (b) and (c) represent the bond energies for $V=0.0t_1, 1.5t_1$ and $-1.5t_1$, respectively, with varying $t_p$. The $B_r$ dominates over $B_{a,b}$ after different $t_p$ values for different values of $V$, implying the onset of a trivial RMI phase with different critical transition points. The green dotted lines in all the plots mark the critical point corresponding to $V=0.0t_1$ in Fig.~\ref{fig:oppPD}.}
 \label{fig:corrs}
\end{figure}

The underlying mechanism of this interaction-induced topological phase transition can be described by the arguments based on the minimization of the ground state energy. Starting from the non-interacting ($V=0$) case (e.g.,  Fig.~\ref{fig:BO2}(a)), the dimerization pattern reveals a large probability of finding two particles on a single rung. At finite $V$, this configuration is unfavourable due to an extra energy cost. In such a situation, to minimize the energy, the system tends to dimerize along the rungs by localizing one particle at each rung. This preference of dimerization along the rungs by the particles leads to a transition from the BO phase (topological) to the RMI phase (trivial) as a function of $V$. This also suggests that a finite $V>0$ favors a topological phase transition and therefore, the transition should occur at a smaller $t_p$ as compared to the $V=0$ case. This is confirmed in our numerical simulations as shown in Fig.~\ref{fig:corrs}, where we compare the average bond energies along the legs ($B_a$ and $B_b$) and the rungs ($B_r$) as a function of $t_p$ for different values of $V$. It can be clearly seen that for $V=1.5t_1$ (Fig.~\ref{fig:corrs}(a)) the bond energy along the rungs starts to dominate over the bond energies along the legs at a smaller $t_p$, indicating an early BO-RMI transition compared to the case when $V=0.0t_1$ (see Fig.~\ref{fig:corrs}(b)).

\begin{figure}[t!]
\includegraphics[width=1\columnwidth]{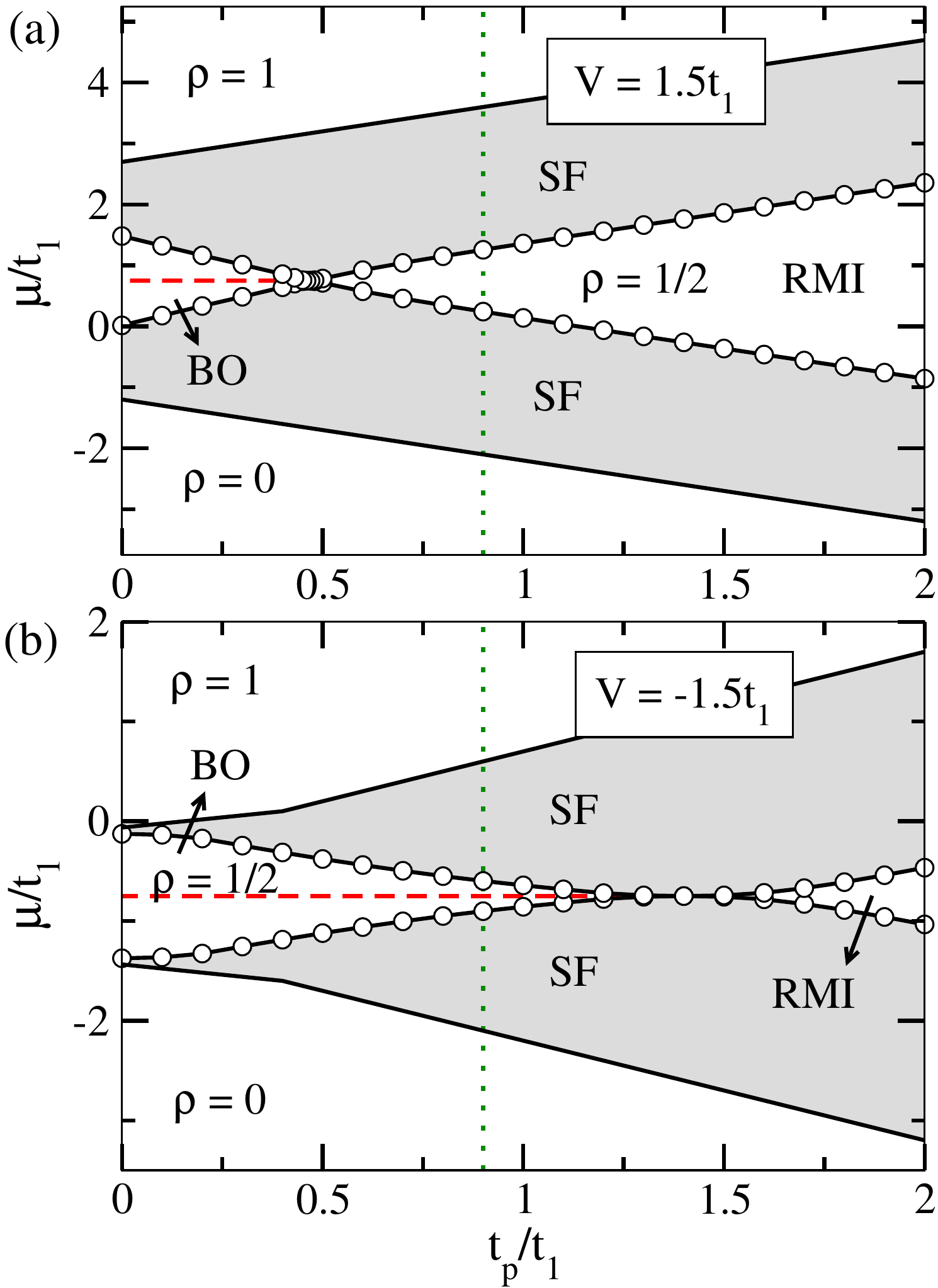}
 \caption{Phase diagrams of Model(\ref{eq:ham}) with (a) $V=1.5t_1$ and (b) $V=-1.5t_1$ are plotted in $t_p$-$\mu$ plane. The solid black lines show the phase boundaries of the gapped topological bond order (BO) phase, and trivial rung-Mott insulator (RMI) phases. The dashed lines stand for the mid-gap edge states. The gapless superfluid (SF) phase is represented by the shaded grey area. The green dotted line marks the critical point corresponding to $V=0.0t_1$ in Fig.~\ref{fig:oppPD}. The change in the gapless critical points for finite interaction can be seen by  comparing to the non-interacting case at half-filling.}
 \label{fig:oppPDv1.5-1.5}
\end{figure}

In this context, we also examine the effect of attractive interaction on the topological phase transition.  We find that the $B_r$ dominates over all other bond-energies at a larger value of $t_p$ for attractive $V$ as shown in Fig.~\ref{fig:corrs}(c) (for $V=-1.5t_1$). This opposite behaviour can be understood using an argument similar to the repulsive $V$ case. When $V<0.0t_1$, the particles prefer to dimerize along the legs rather than on the rungs. Hence, a larger $t_p$ is necessary to break the bond ordering along the legs and introduce rung dimerization as compared to the repulsive $V$ case. These behaviors can also be seen in the phase diagrams shown in Fig.~\ref{fig:oppPDv1.5-1.5} (a) and (b) for $V=1.5t_1$ and $V=-1.5t_1$ respectively. Comparing the transition point for $V=0.0t_1$ marked by vertical dotted lines, we can clearly see that, the topological phase transition occurs at a smaller (larger) critical $t_p$ for repulsive (attractive) $V$.

\begin{figure}[t]
\includegraphics[width=1\columnwidth]{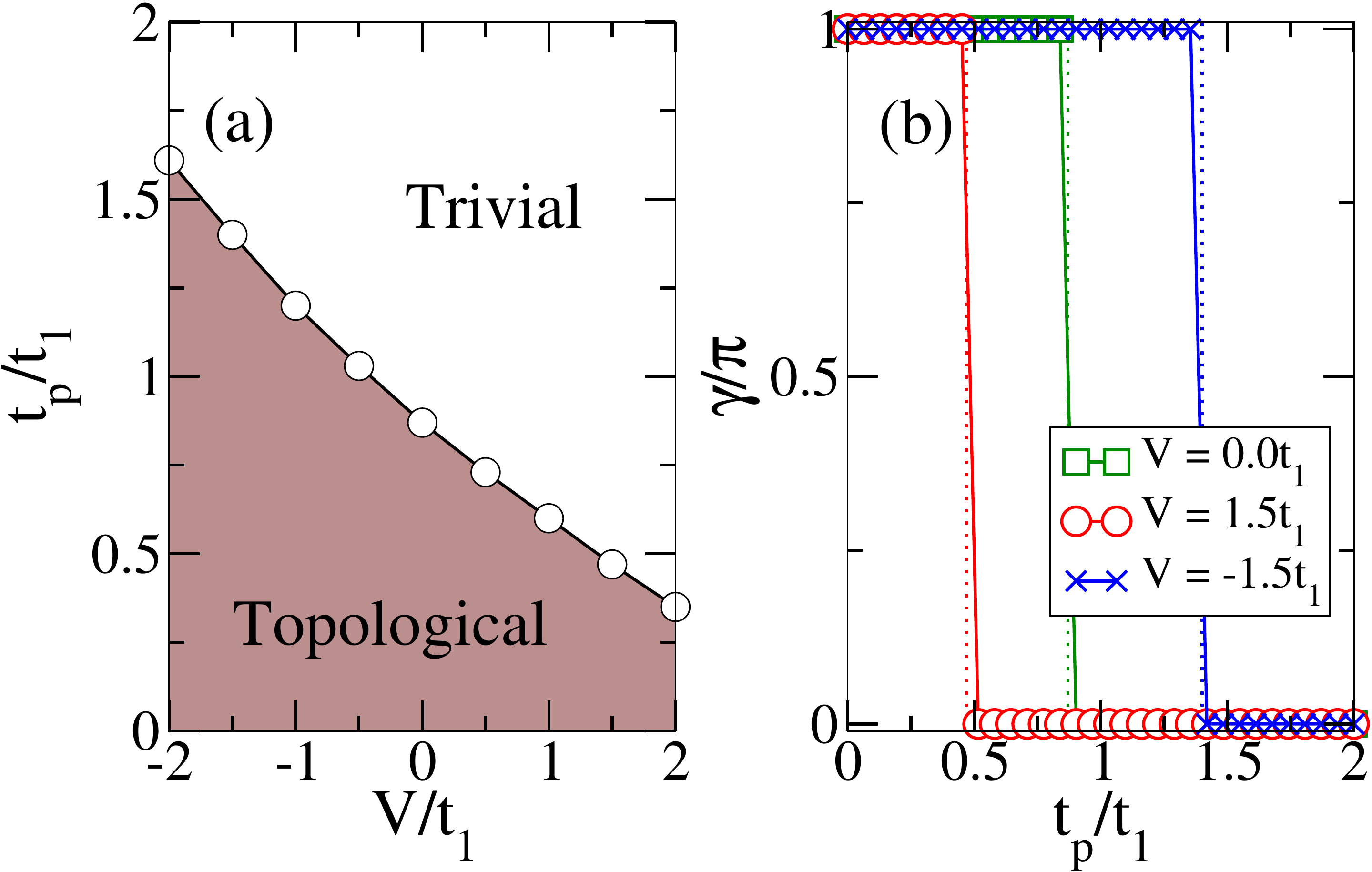}
 \caption{(a) Phase diagram at $\rho=1/2$ corresponding to the Hamiltonian given in Eq.~\ref{eq:ham} for $t_a= t_b = 0.6t_1$ and $\alpha_a = -\alpha_b = 0.4t_1$ (staggered dimerization). Here the topological (trivial) phase is the BO (RMI) phase. The figure shows how the critical point changes with the interaction strength $V$. (b) Berry phase under twisted phase boundary conditions showing the topological phase transition as a function of $t_p/t_1$ for $V=0.0t_1, 1.5t_1$ and $-1.5t_1$ on a system of length $L=6$ ($12$ lattice sites). The dotted lines mark the transition points extracted from the phase diagram in (a) corresponding to each $V$.}
 \label{fig:PDV}
\end{figure}

To  quantify the interaction induced topological phase transition, we obtain the critical rung hopping strengths $t_p^c$ for different values of $V$ by monitoring the bulk gap-closing point (as already done for Fig.~\ref{fig:oppPD} and Fig.~\ref{fig:oppPDv1.5-1.5}). To this end we fix $t_a= t_b = 0.6t_1$ and $\alpha_a = -\alpha_b = 0.4t_1$ in the Hamiltonian shown in Eq.~\ref{eq:ham} and compute the phase diagram in the $V-t_p$ plane as shown in Fig.~\ref{fig:PDV} (a) where the topological phase (brown region)  and trivial phase (white region) are separated by the critical boundary (line with circles) . We obtain that the $t_p^c$ shifts to a higher (lower) value for attractive (repulsive) interaction.

The topological phase transitions can also be detected from the discontinuous jump in the topological invariant of the corresponding phases. A bulk topological invariant, calculated from the Berry phase, is always linked with the symmetry-protected edge states according to bulk-boundary correspondence in topological systems. The Berry phase, defined by the formula
\begin{equation}
\gamma = \int_{0}^{2\pi}\langle\psi(\theta)|\partial_{\theta}\psi(\theta)\rangle d\theta
\end{equation}
can be a suitable topological invariant for our model in the many-body limit under twisted phase boundary conditions (TBCs)~\cite{Grusdt2013}. Here $|\psi\rangle$ is the ground state wavefunction, and we achieve the TBC by setting $a_j\to e^{i\theta/L}a_j$ and $b_j\to e^{i\theta/L}b_j$ in leg-a and leg-b, respectively. When the twist angle $\theta$ is varied from $0$ to $2\pi$ adiabatically, $|\psi\rangle$ picks up a phase, which is nothing but the Berry phase. Thus, $\gamma$ is expected to be quantized in units of $\pi$ for a topological phase, whereas it should vanish in the trivial phase. We plot $\gamma/\pi$ as a function of $t_p$ in Fig.~\ref{fig:PDV} (b) to capture the topological phase transitions for $V=0.0t_1, 1.5t_1$ and $-1.5t_1$. As anticipated, $\gamma$ clearly distinguishes the topological BO and trivial RMI phases for all three values of $V$. It also marks the respective critical points ($\sim 0.87t_1$ for $V=0.0t_1$, $\sim0.47t_1$ for $V=1.5t_1$ and $\sim1.4t_1$ for $V=-1.5t_1$), where we see an abrupt jump from $\gamma=\pi$ to $\gamma=0$.

\subsection{Thouless charge pumping} \label{sec:pump}

We now propose an experimentally relevant quantity in terms of Thouless charge pumping (TCP)\cite{Thouless1983} which can detect the interaction-induced topological phase transition arising from our model. The TCP has recently been used to characterize the topological nature of a system both in theory as well as in experiments~\cite{Citro2023, Lohse2016,Takahashi2016pumping, Schweizer2016, Kraus2012pumping, Taddia2017, walter2022quantisation, dai_pump}. 
As per the TCP measure, it is possible to pump a quantized amount of charge  with an adiabatic variation of the system parameters in a pumping cycle which is related to the Chern number. The celebrated Rice-Mele (RM) model~\cite{Rice1982, Asboth2016_rm} defines the pumping protocol in one dimension where in the parameter space, the pumping path winds around a gapless singular point~\cite{Hayward2018}. Here the system periodically and adiabatically goes from non-trivial to trivial and to non-trivial phase again by breaking the symmetry that protects the topology in the system. While originally, the RM model described the TCP of non-interacting systems, recently its connection to interacting systems has been proposed in various systems~\cite{Mondal2020, mondal_sshhubbard,Hayward2018,Schweizer2016, Nakagawa2018, Berg2011, bertok_pump, mondal_phonon, Meden, kuno_pump}. For our current system under consideration, which exhibits a topological phase transition, we can define a pumping protocol by introducing a symmetry breaking term such that the pumping path in the parameter space adiabatically winds around the topological phase transition point. 

\begin{figure}[!t]
\centering
\includegraphics[width=1\columnwidth]{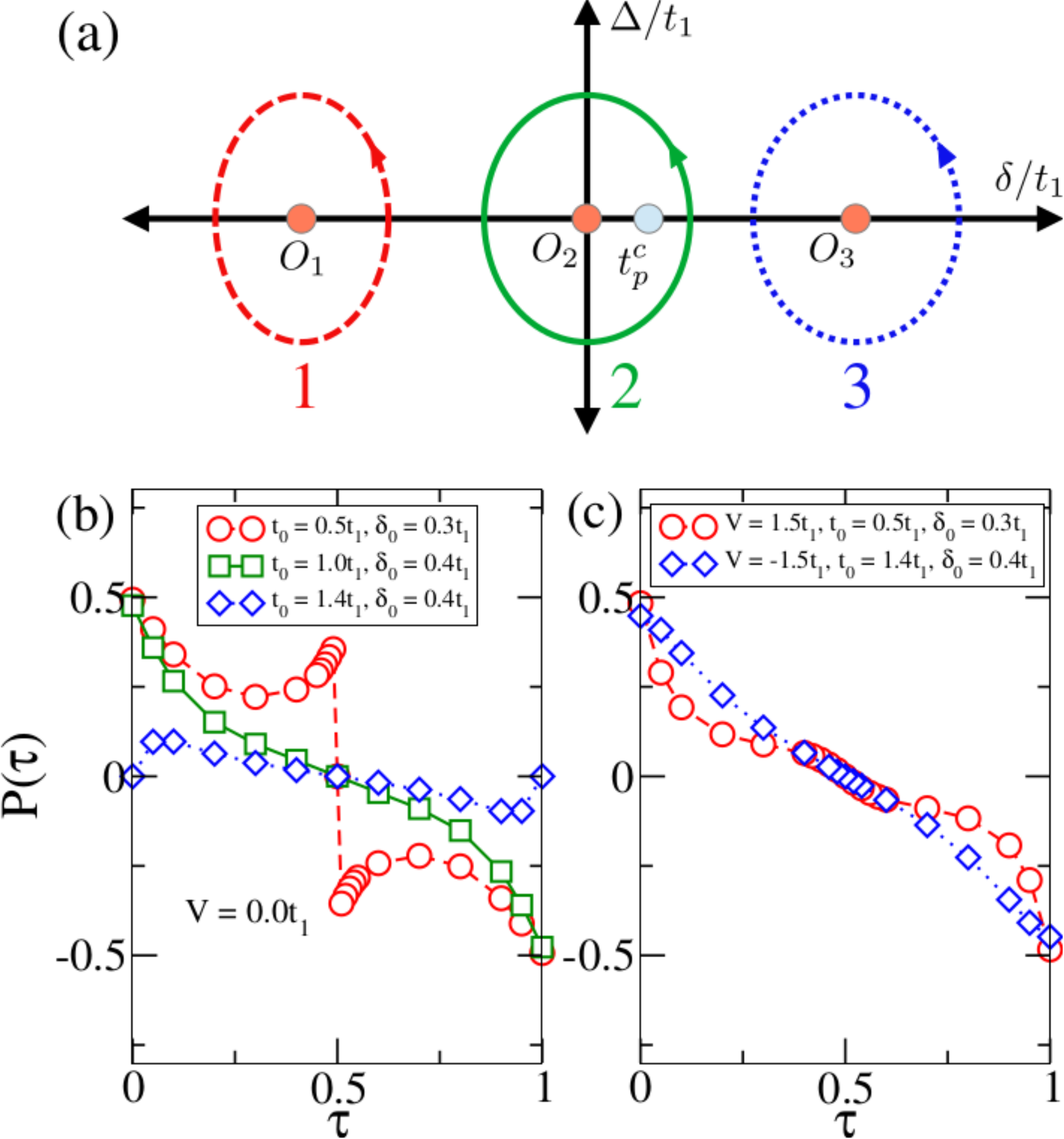}
 \caption{(a) Pictorial representation of the adiabatic variation of parameters are shown for three different pumping cycles with three different origins ($t_o$'s). The topological phase transition critical point $t_p^c$ is marked with a green circle on the $\delta$-axis. (b) The evolution of the polarization ($P(\tau)$) is plotted for three different $t_o$'s corresponding to three different pumping cycles shown in (a) for $V=0$. Here $t_p^c \sim 0.87t_1$. Here only cycle-2, which encloses the $t_p^c$, shows robust pumping. (c) The evolution of $P(\tau)$ is shown for the same parameters corresponding to cycle-1 and 3 that are considered in (b), but with finite interaction strength $V=1.5t_1$ and $V=-1.5t_1$ respectively. Unlike the non-interacting case, here a quantized amount of charge is being pumped. We call this phenomenon as the  interaction-induced topological charge pumping. Here, for all the cases we consider $\Delta = 0.5t_1$ and a finite system of $L=200$ rungs.}
 \label{fig:pump}
\end{figure}

In the following, we present the pumping protocol for the two-leg ladder system with the parametric extension of the model~\ref{eq:ham} which is given as
\begin{align}
\mathcal{H}_{p}(\tau) = &-\sum_{j} (t_{a}-(-1)^{j}\alpha_a)  (a_{j}^{\dagger}a_{j+1}^{\phantom \dagger} + \text{H.c.}) \nonumber\\
&-\sum_{j} (t_{b}-(-1)^{j}\alpha_b)  (b_{j}^{\dagger}b_{j+1}^{\phantom \dagger} + \text{H.c.}) \nonumber\\
&- (t_o + \delta(\tau))\sum_{j}(a_{j}^{\dagger}b_{j}^{\phantom \dagger}+\text{H.c.}) \nonumber\\
&+ \Delta(\tau) \sum_j \left((-1)^j n_{aj} + (-1)^{j+1} n_{bj}\right) \nonumber\\
&+ V\sum_{j} n_{aj}n_{bj},
\label{eq:hamP}
\end{align}
where $\delta(\tau) = A_\delta\cos(2\pi \tau)$ changes the hopping dimerization along the rungs and $\Delta(\tau) = A_\Delta\sin(2\pi \tau)$ changes the staggered onsite potential, which breaks the sublattice symmetry. The quantity $\tau$ is the adiabatic pumping parameter with $O=(t_p = t_o,~0)$ as the origin of the pumping cycle in the $\delta-\Delta$ plane. A schematic representation of the periodic variation of the parameters is shown in  Fig.~\ref{fig:pump}(a) for three different $t_o$'s with $A_\delta$ and $A_\Delta >0 $. Note that the pumping can only happen if the pumping path encloses the gap-closing critical point ($t_p^c$). This implies that only for the cycle-$2$ (green continuous line) of Fig.~\ref{fig:pump}(a) we can expect robust and quantized charge pumping.

However, the results obtained for $V=0$ suggest that the critical rung hopping $t_p^c$ for the topological phase transition can be moved towards smaller and larger values  depending on the repulsive or attractive nature of $V$, respectively. Hence, it can be made possible to shift the $t_p^c$ along the $\delta$ axis of Fig.~\ref{fig:pump}(a) by a suitable choice of $V$ such that it lies inside the cycle-1 or 3 and a finite quantized amount of pumped charge can be generated in these cases. The phenomenon of charge pumping induced by interactions, can be termed as an interaction-induced topological charge pumping  (iTCP).

In our analysis, the iTCP is demonstrated by defining three different pumping cycles for three parameter sets of ($t_o,~A_\delta,~A_\Delta$) in Fig.~\ref{fig:pump}(a) such as  cycle-1: $(0.5t_1,~0.3t_1,~0.5t_1)$, cycle-2: $(1.0t_1,~0.4t_1,~0.5t_1)$ and cycle-3: $(1.4t_1,~0.4t_1,~0.5t_1)$. All the cycles correspond to the parameter sets $t_a= t_b = 0.6t_1$, $\alpha_a = -\alpha_b = 0.4t_1$, for which the critical rung hopping for the non-interacting ($V=0$) topological phase transition is $t_p^c \sim 0.87t_1$ (Fig.~\ref{fig:oppPD}) at half-filling. Following the standard protocol of the charge pump, we compute a quantity known as polarization using the formula
\begin{equation}
 P(\tau) = \frac{1}{L} \sum_{j=0}^{L-1} \langle \psi(\tau) | (j-L/2) n_{j} | \psi(\tau) \rangle
\end{equation}
which can estimate the pumped charge ($Q$) as,
\begin{equation}
 Q = \int_0^1 d\tau \partial_\tau P(\tau).
\end{equation}
Here, $| \psi(\tau) \rangle$ is the ground state corresponding to the Hamiltonian $\mathcal{H}_{p}(\tau)$ since the evolution is adiabatic. In Fig.~\ref{fig:pump}(b), we plot $P(\tau)$ for $V=0$, for the pumping cycle-1 (dashed line with circles), cycle-2 (continuous line with squares) and cycle-3 (dotted line with diamonds). According to the parameters considered, the $t_p^c$ resides on the $\delta$ axis within the cycle-2 ($t_o-A_\delta < 0.87t_1 < t_o+A_\delta$) only. Thus, we can see from the figure that only for cycle-2, $P(\tau)$ varies continuously from $0.5$ to $-0.5$ resulting in $|Q| = 1$. However, for cycle-1 and cycle-3, the situation is completely different. For cycle-1, there is a clear breakdown of charge pumping in the middle of the pumping cycle, and there is no charge being pumped for cycle-3.

To verify the iTCP, we re-perform the pumping along cycle-1 and 3 with finite interactions $V=1.5t_1$ and $-1.5t_1$, which shift the $t_p^c$ into cycle-1 and cycle-3, respectively. In Fig.~\ref{fig:pump}(c), we plot  $P(\tau)$ for cycle-1 (dashed line with circles) and cycle-3 (dotted line with diamonds). The continuous change of $P(\tau)$ from $0.5$ to $-0.5$ clearly shows a quantized TCP for both the cases. Note that this finite TCP was not present for the $V=0$ case (Fig.~\ref{fig:pump}(b)), and it is induced by the interaction ($V$), indicating an iTCP. This analysis substantiates the interaction-induced topological phase transition as already shown in Fig~\ref{fig:PDV}. 

From the analysis above, we can see that interaction favours a topological phase transition from topological BO to the trivial RMI phase. The same underlying physics also dictates the change in the critical rung hopping of topological phase transition in the presence of interaction. Such an interaction effect on the topological phase transition allows for an interaction-induced topological charge pumping.

\section{Summary and Outlook}
\label{sec:conc}
In summary, we have studied the topological properties of the bosonic Su-Schrieffer-Heeger ladder for its two possible configurations. These configurations correspond to the uniform dimerization case where the dimerization pattern is such that both the legs are topological for the appropriate choice of boundary conditions, and the staggered dimerization case, where one of the legs is topological and the other is trivial. Compared to previous studies of either single particle physics or fermions, we analyzed hardcore bosons hopping on the ladder and endowed with interleg interactions.   We first analyzed the ground state properties of the many-body system without any inter-leg interaction and showed that for the uniform dimerization case, there is no topological phase transition as a function of the rung hopping strength. Rather, the system goes from a gapped bond order phase to a gapped rung-Mott insulator phase at half-filling without any gap-closing. In contrast, the staggered dimerization case supports a well defined topological phase transition from a topological bond order phase to a trivial rung-Mott insulator phase. Further, we investigated the effect of inter-leg interaction which greatly influences the topological phases. We found a topological phase transition as a function of interaction for a fixed rung hopping strength. We also found that when interaction is fixed, the critical rung hopping for the topological phase transition strongly depends on the nature of the interaction. That is, for repulsive (attractive) interaction the topological phase transition occurs at a smaller (larger) critical rung hopping strength. We characterized the transitions through multiple approaches and density matrix renormalization group (DMRG) techniques; these included obtaining the excitation gap, edge state profiles, Berry phases and Thouless charge pumping measures. 

As these studies are the first to be performed--to the best of our knowledge,--on hard core bosons hopping on the SSH ladder and under the influence of inter-leg interactions, there is scope for extensive further studies. To name a few, the phase diagram warrants a more careful characterization. A highlight feature of bosons is their ability to host superfluidity; further studies could deviate from half-filling to study the entry from the superfluid phase into the topological insulating phases. While DMRG provides a tractable technique for deriving properties of the phase diagram, alternative ways of incorporating the effects of interaction, such as Luttinger liquid treatment, would shed light on the nature of the phase transitions and the relevance of various coupling terms.  As for the derived measures, Thouless charge pumping is new in the context of interacting bosons and warrants further analysis. Lastly, generalizations of this bosonic SSH ladder model would include deviations from the hardcore constraint to finite on-site interaction strength, multiple legs of the ladder, inhomogeneity stemming from a spatially varying site-potential, and frustration due to modified lattice structure. 

Along with more extensive theoretical study, the bosonic SSH ladder system is poised to be explored in experiment. Several ultracold atomic systems have characterized single bosonic SSH chains and topological phases in optical lattice systems in real space as well as in momentum space~\cite{gadway}, employing time-of-flight~\cite{Atala2013}, mean chiral displacement~\cite{Xie2019}, and Thouless pumping measures~\cite{Takahashi2016pumping, Lohse2015} for identifying topological features. These avenues would be amenable to studies of interacting ladders, with topological properties directly measurable, for instance, through Rydberg atoms~\cite{browyes}.
Mechanical systems~\cite{prodan} have also realized SSH ladders in regimes where simple harmonic approximations hold; going beyond could introduce the equivalent of interactions in the relevant degrees of freedom. Wave-guide arrays, photonic crystals, and superconducting Josephson circuits have all come far in the context of lattice-based strongly correlated physics, and would be highly amenable to the system at hand. The SSH bosonic ladder would therefore offer rich ground for theory and experiment to go hand-in-hand.


\section{Acknowledgment} 
We dedicate this work in memory of Prof. Amit Dutta who was a great physicist, teacher, and mentor to many and who has been a cherished presence to us as a fellow-scientist and friend. TM acknowledges  support from Science and Engineering Research Board (SERB), Govt. of India, through project No. MTR/2022/000382 and STR/2022/000023.

\bibliography{references}

\end{document}